\documentstyle[12pt,vassilis_moriond]{article}
\begin{document}
\heading{The Fate of Ultra--Luminous Mergers}

\author{A. C. Baker $^{1}$, D. L. Clements $^{2}$} {$^{1}$ Service
d'Astrophysique, Saclay, France \\$^{2}$ Institute d'Astrophysique
Spatiale, Orsay, France}{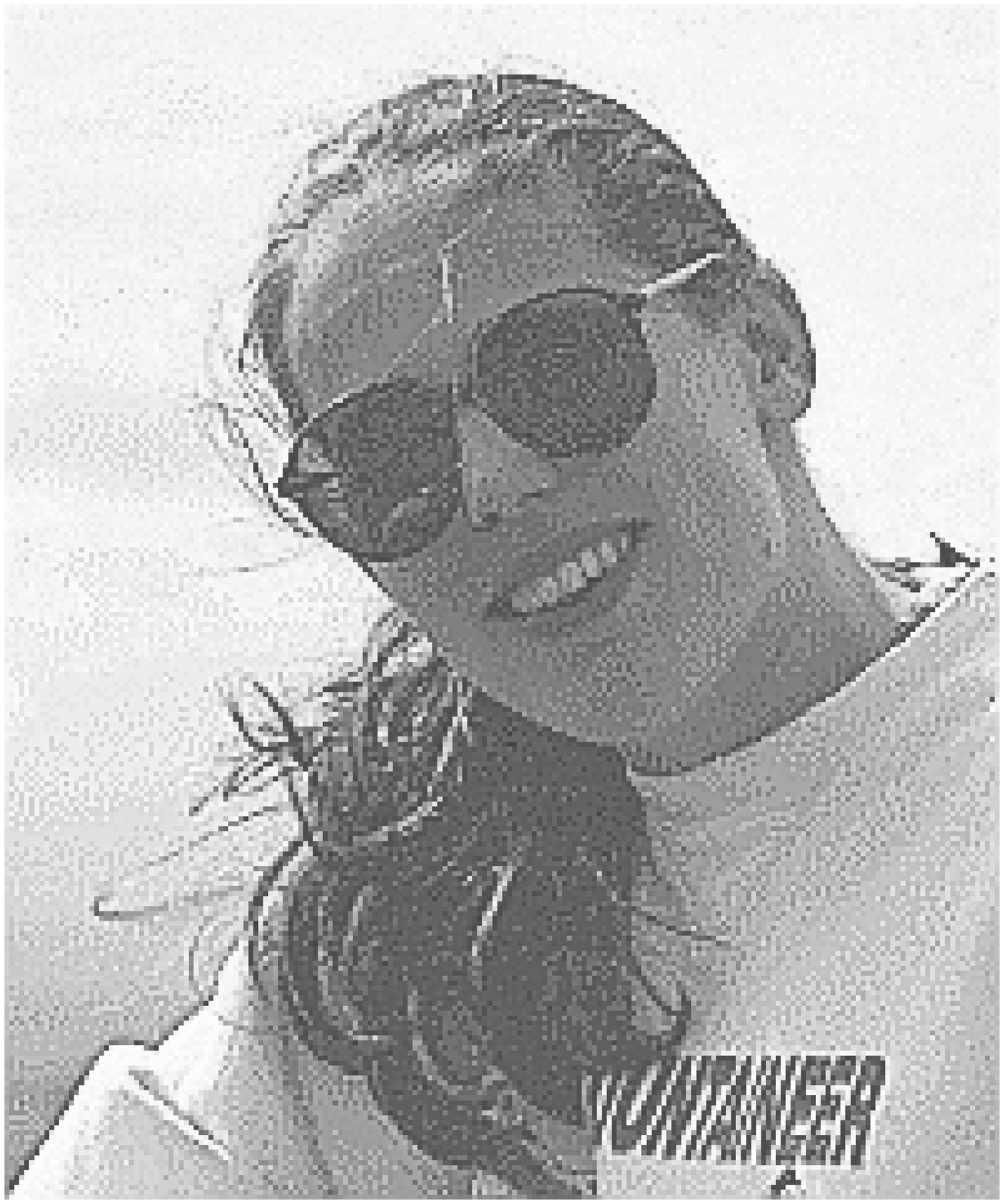}

\begin{moriondabstract}

Essentially all Ultra-Luminous IR Galaxies (ULIRGs) are in disturbed,
interacting or merging systems.  It is known that interactions tend to
induce galactic starbursts. Thus, elliptical galaxies which are formed
in mergers will tend to have high metallicity, low dust and molecular
gas content, and faint structural distortions, as observed for the
{\it bone fide} elliptical galaxy population. The old stellar
population in the merged galaxies will probe the new gravitational
potential, relaxing rapidly to give a de Vaucouleurs
surface--brightness profile if the remnant is elliptical--like.  We
examine the old stellar population in 10 nearby\footnote{In this
paper, we assume $q_0 = \frac{1}{2}$ and $H_0$ as stated.} (z $ < 0.15
$) ULIRGs, using deep near--IR imaging photometry. These data reveal
signs of elliptical--like structure in the near--IR one--dimensional
surface brightness profiles, supporting the hypothesis that
ultra--luminous mergers evolve into elliptical galaxies.

\end{moriondabstract}

\section{Motivation : Mergers and Galaxy Formation}

The fundamental question underpinning this study is : `What
astrophysical processes determine whether a galaxy forms as a disk or
an elliptical system?' It is now clear that galaxy formation is not an
unique event which occurs in the early universe, but rather, a diverse
range of processes which are ongoing even to the present
epoch. Numerical simulations~\cite{Barnes89} suggest that collisions
between disk galaxies form systems which pass through a massive burst
of star formation, and eventually produce `elliptical--like' remnants
with mass distributions similar to those inferred for {\it bone fide}
elliptical galaxies. Toomre~\cite{Toomre77} showed that the observed
relative numbers of merging systems and elliptical galaxies are
consistent with a significant fraction of ellipticals forming {\em
via} the merging of spiral galaxies, an ironic inversion of Hubble's
original sequence. Theories of galaxy formation divide into
`top--down' and `bottom--up' scenarios. The merging of smaller disk
galaxies to form massive spheroidal galaxies fits well with the
bottom--up `hierarchical clustering' galaxy formation schemes
envisaged in cosmologies where cold, dark mass dominates the
Universe. It is therefore plausible that some ultra--luminous mergers
are `factories' turning molecular gas--rich, dusty disk galaxies into
elliptical--like systems. In order to advance an observational study
of the r\^{o}le of disk--disk mergers in the formation of elliptical
galaxies, a sample of galaxy systems in the process of merging is
needed.

Observationally, luminous far--IR (FIR) emission is known to trace
star formation \cite{Sanders96}.  The {\em ultra}--luminous IR
galaxies (ULIRGs) are almost all disturbed, interacting or merging
systems (Clements and co--workers \cite{Clements96a,Clements96b}).
ULIRGs are distinguished by $ L_{\mathrm{bol}} \sim L_{\mathrm {FIR}}
= 10^{12} L_{\odot} $ ($H_0 = 75 {\mathrm{km s^{-1} Mpc^{-1}}} $),
which means that they are amongst the most luminous galaxies in the
Universe. The prodigious far--infrared (FIR) emission from ULIRGs
appears to be triggered by the interaction and merger of galaxies, and
produced by very warm dust.  It seems clear that for the vast majority
of the population of luminous IR galaxies, the dust is heated by
emission from regions of star formation. However, in the most luminous
systems, there is often also evidence of an active nucleus, producing
Seyfert--like high--ionization emission lines with velocity widths
much greater than those of normal
galaxies~\cite{Sanders96,Clements96a}. An excellent review of luminous
IR galaxies has been given by Sanders and Mirabel~\cite{Sanders96}.

It is clear that ULIRGs represent an easily--identifiable sample of 
galaxy mergers, and are ideal for the present purpose.

\section{The Infrared Imaging Data}

For this study, we selected ULIRGs from the
literature~\cite{Leech94,Zhenglong91,Melnick90} by two main
criteria. We required targets which are sufficiently nearby (redshifts
$ z < 0.15 $) for good spatial resolution, and which are all confirmed
as interacting and merging systems by previous observations. The basic
characteristics of the sample are given in Table~\ref{t:sample}. One
ULIRG in our sample, IRAS 16487+5447 had been classified as `isolated
and undisturbed' based upon optical imaging~\cite{Leech94}. However,
our near-IR data revealed an unambiguous double nucleus. The evidence
that essentially all ULIRGs are in disturbed, interacting or merging
systems is now becoming overwhelming~\cite{Sanders96, Clements96b}.

The aim of the current project is to use the old stellar population as
test particles to probe the gravitational potential resulting from
galaxy interaction and merger. The integrated emission contributed by
the pre-existing old, red stars, relative to the young, blue burst
stars, is much greater in the near--IR than at optical
wavelengths. Therefore, we obtained deep K--band images of the 10
ULIRGs using the Blue MAGIC infrared camera on the Calar Alto 3.5m
telescope.  Our data have a spatial scale of $ 0.32^{''} /
\mathrm{pixel} $, so the field--of--view is $ 82^{''} \times 82^{''}$,
which is roughly equivalent to 10 -- 20 effective radii
($r_{\mathrm{eff}}$) for the target galaxies. The exact physical scale
is given by the standard formula \cite{Lang86} $ l = l(z,\phi,
h_{100}) = 29.07 \frac{\phi^{''}}{h_{100}} f(z) $ where $ f(z) =
\left( 1+z \right)^{-2} \left[ \left( 1+z \right) - \left( 1+z
\right)^{\frac{1}{2}} \right] $ depending upon the redshift $z$, the
angular separation $\phi$ in arcseconds and the Hubble parameter $H_0
= 100\, h_{100}\, \mathrm{km s^{-1} Mpc^{-1}}$.

We used the narrower K--band filter, Km (based on that described by
Wainscoat and Cowie~\cite{Wainscoat91}), since this reduces the
contribution of thermal sky emission to the background flux. The
exposures were taken in a dithered pattern, with total integration
times in the range 1000 -- 2700 s per ULIRG. Therefore, the
sensitivity of our data reaches $ \sim 20.5 \; \mathrm{mag} \;
\mathrm{arcsec}^{-1}$. The seeing was generally good, being typically
$ < 1^{''}$, with a few poorer intervals.

The data were reduced using the IRAF packages\footnote{IRAF is
distributed by the NOAO, which are operated by the Association of
Universities for Research in Astronomy, Inc., under cooperative
agreement with the National Science Foundation.} and standard
techniques. We made measurements of photometric reference stars
throughout the observing run. However, there were problems with
absolute photometric calibration due to the defocusing used to prevent
the standards from saturating the MAGIC detector. The photometric
calibrations have been verified by comparison with previous published
aperture photometry of Arp 220
\cite{Carico90}.

\section{Surface Brightness Profiles}

One straight--forward discriminator between elliptical and disk
galaxies is the shape of the azimuthally--averaged surface brightness
profile. As a first step in determining the fate of the
ultra--luminous mergers in our sample, we have examined these
one--dimensional surface--brightness profiles. The profiles were
calculated by averaging the flux in elliptical annuli using the IRAF
{\bf ellipse} routine. We then fit analytic surface brightness
profiles of the form

$$ 
\propto \exp \left [ - \left ( \frac{r}{r_s} \right )^{\beta} \right ]
			\left\{ \begin{array}{ll} 
				\beta = 1 	& \mbox{disk} \\
				\beta = 0.25 	& \mbox{elliptical} 
			\end{array}
		\right . $$

The shape parameter $\beta$ takes a value $\beta = 1$ in disk galaxies
(the exponential disk profile), and $\beta = \frac{1}{4}$ in
elliptical galaxies (the de Vaucouleurs $r^{\frac{1}{4}}$ law).  In
principle, the value of $\beta$ could depart from the shape of these
traditional spiral and elliptical galaxy structures. We therefore
calculated fits in which $\beta$ was unconstrained, but these did not
result in a significantly better description of our data.

We use the standard $\chi^2$ `goodness--of--fit' parameter to choose
between the possible descriptions of the distribution of the infrared
light in the ULIRGs in our sample. A potential problem with this
approach is that we are interested in the {\em overall} distribution
of the old stellar population, whereas the $\chi^2$ parameter is
sensitive to all the detailed structure down to the spatial resolution
of the data. Therefore, we do not expect to obtain fits such that
$\chi^2 = 1$ per degree of freedom. Still, a lower value of $\chi^2$
still means that the differences between the data and the model are
reduced. We therefore use the ratio between the values of the $\chi^2$
parameter for the disk and elliptical description of each galaxy to
decide which is our {\it prefered} description in each case.  A more
refined approach could use the colours of each galaxy to identify and
hence mask out of the fit those regions which are not dominated by old
stellar light.

\begin{table}[ht]
\label{t:sample}
\begin{center}
\begin{tabular}{llll}
IRAS name 	& Alternative 	& Redshift$^1$ 	& $\chi^2$ 	\\
of ULIRG	& Name(s)     	& $z$		& ratio		\\ \hline
00015$+$4937	&		& 0.148		& 5.8		\\
15250$+$3609	& PGC 055114	& 0.0554	& 0.92		\\
15327$+$2340	& Arp 220	& 0.01813	& 0.05		\\
16487$+$5447	&		& 0.1044	& 0.52		\\
17179$+$5445	&		& 0.1475	& 0.38		\\
17208$-$0014	& PGC 060189	& 0.04281 	& 0.32		\\
19297$-$0406	&		& 0.08573 	& 0.42		\\
20414$-$1651	&		& 0.08708 	& 0.17		\\
20087$-$0308	&		& 0.10567 	& 0.16		\\
22491$-$1808	& PGC 069877	& 0.07776 	& 0.09	\\ \hline
\end{tabular}
\end{center}

\caption{Information about the Ultra--Luminous IR Galaxies.}

$^1$ {\small Redshift data are taken from the NASA/IPAC Extragalactic
Database (NED), which is operated by the Jet Propulsion Laboratory,
California Institute of Technology, under contract with the National
Aeronautics and Space Administration.}
\end{table}

The results are given in Table~\ref{t:sample}. In 8 of our 10 systems,
we strongly favour the elliptical--like description. For 15327$+$2340,
we cannot discriminate between the two descriptions using our current
approach. For the remaining, ULIRG 00015$+$4937, we clearly favour the
disk--like description. This is a clear double nucleus system, and the
data given are for the brighter of the two nuclei. Judging by the
large separation of the nuclei in this case, the merger is at a
relatively early stage. We are therefore detecting the pre-existing
galaxy structure of what is probably a molecular gas--rich spiral.

\section{Arp 220}

As an oft--cited example of a `prototypical' ULIRG, it is to be
expected that the results for Arp 220 will be unusual in any study of
the ULIRG population. The fierce debate over the energy budget of Arp
220 is ongoing. New evidence has emerged to support the hypothesis
that the bulk of the FIR emission is fuelled by a massive
starburst~\cite{Lutz96}. However, the hypothesis that a hidden AGN is
the dominant power--source is also well--supported. For example, new
mid--IR spectrophotometry indicates that the heating source for the
dust is smaller than a few parsecs~\cite{Dudley97} . It seems
increasingly likely that Arp 220 is a hybrid system, with significant
energy contributions from both an active nucleus and star formation
activity.

\begin{figure}[!ht]
\label{f:arp220}
\psfig{figure=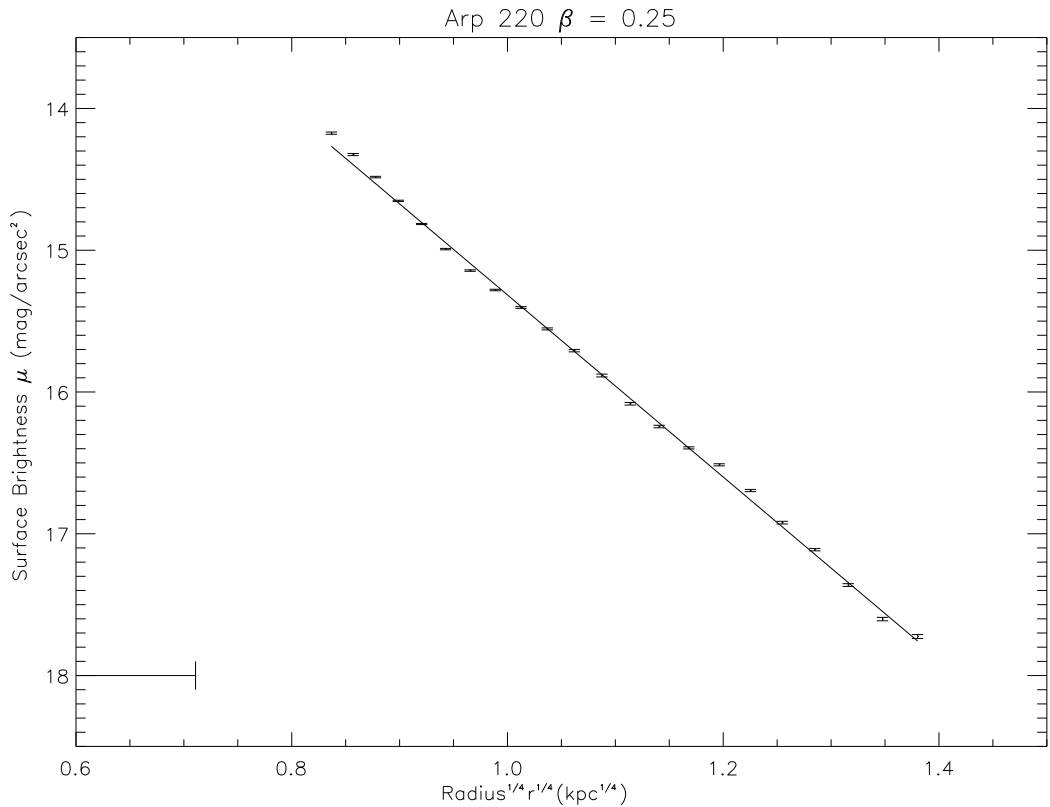,angle=-270,width=15cm}
\caption{The Surface Brightness Profile of Arp 220}
\end{figure}

Previous work on Arp 220 by Wright et al. \cite{Wright90} used data
taken in relatively poor seeing ($\sim 2^{''}$) and with a spatial
resolution ($0.62^{''}/\mathrm{pixel}$), half that of the present
study. They determined that Arp 220 is an elliptical--like system
($\beta = \frac{1}{4}$) with an effective radius $r_{\mathrm{eff}} =
3.8 \mathrm{kpc}$. Our new results confirm that the overall structure
of Arp 220 is elliptical--like, although we find a somewhat smaller
effective radius of $r_{\mathrm{eff}} = 2.4 \mathrm{kpc}$, consistent
with our better seeing and spatial resolution. The data and model for
the surface brightness profile of Arp 220 are shown in
Figure~\ref{f:arp220}.

\section{The Fate of Ultra--Luminous Mergers}

The one--dimensional surface--brightness profiles of these ULIRGs show
a strong tendancy to be better described by a de Vaucouleurs
$r^{\frac{1}{4}}$ law than by an exponential disk. This is the case
for 8 of the 10 systems in the present study. We infer that
ultra--luminous mergers can process (disk) systems which are rich in
molecular gas, eventually relaxing into (molecular gas--depleted)
elliptical--like systems. This is yet another indication that many
`normal' galaxies are far from serene. Our results support the
suggestion that the collision and merger of classical spiral galaxies
can produce classical elliptical galaxies, through astrophysical
processes such as tidal disruption, violent star formation accompanied
by prodigious infrared emission, and gravitational relaxation.

\acknowledgements{The data were obtained at the German--Spanish
Astronomical Centre, Calar Alto, jointly operated by the
Max--Planck--Institute for Astronomy, Heidelberg, and the Spanish
National Commission for Astronomy. We would like to thank Tom Herbst
for his assistance in obtaining these data. This work was supported by
the EC TMR Network programme, FMRX-CT96-0068. ACB thanks the European
Southern Observatory for hospitality during the data analysis.}


%
\vfill
\end{document}